\documentclass[usenatbib,onecolumn]{mn2e}
\usepackage{amsfonts}
\usepackage{graphicx}
\usepackage{newtxtext,newtxmath}
 
\usepackage{lipsum}
\usepackage[T1]{fontenc}
\usepackage{ae,aecompl}
  \usepackage{array}
  \usepackage{hyperref}
  \usepackage{pdflscape}
\usepackage{graphicx}   
\usepackage{amsmath}    
\usepackage{amssymb}    
\usepackage{bm} 
\usepackage{multirow}
\usepackage{multicol}
\usepackage{cuted}
\usepackage{mathtools}
\def \aj {{\rm {AJ}}}

\def \apj {{\rm {ApJ}}}
\def \icarus {{\rm {Icarus}}}
\def \procspie {{\rm {Proc. SPIE}}}
\def \apjs {{\rm {ApJS}}}

\def \aap {{\rm {A\&A}}}
\def \pasp{\rm {PASP}}

\def \aaps {{\rm {A\&AS}}}

\def \mnras {{\rm {MNRAS}}}

\def \apjl{\rm {ApJL}}
\def \nat{\rm {Nat.}}

\def \mj{\,$M_{\rm Jup}$\,}
\def \msun{\,$M_\odot$\,}

\def \masyr{\,mas\,yr$^{-1}$}

\date{\today}

\title[Detection of the smallest cold Jupiter]{Optimized modeling of Gaia-Hipparcos astrometry for the detection of the smallest cold Jupiter and confirmation of seven low mass companions}
\author[F. Feng et al.]
{Fabo Feng$^{1,2}$\thanks{E-mail: ffeng@sjtu.edu.cn},
  R. Paul Butler$^{2}$, Hugh R. A. Jones$^{3}$, Mark
  W. Phillips$^{4}$, Steven S. Vogt$^{5}$,
  \newauthor
 Rebecca Oppenheimer$^{6}$, Bradford
  Holden$^{5}$, Jennifer Burt$^{7}$, Alan P. Boss$^{2}$\\
$^{1}$Tsung-Dao Lee Institute, Shanghai Jiao Tong University, 800 Dongchuan Road, Shanghai 200240, People's Republic of China\\
$^{2}$Earth and Planets Laboratory, Carnegie Institution for Science, Washington, 5241 Broad Branch Road, NW, Washington, DC 20015, USA\\
$^{3}$Centre for Astrophysics Research, University of Hertfordshire, College Lane, AL10 9AB, Hatfield, UK\\
$^{4}$Astrophysics Group, University of Exeter, EX4 4QL, Exeter, UK\\
$^{5}$UCO/Lick Observatory, University of California, Santa Cruz, CA 95064,USA\\
$^{6}$Astrophysics Department, American Museum of Natural History, Central Park West at 79th Street, New York, NY 10024, USA\\
$^{7}$Jet Propulsion Laboratory, California Institute of
  Technology, 4800 Oak Grove drive, Pasadena CA 91109\\
}
\date{\today}

\begin{document}
\maketitle
\begin{abstract}
To fully constrain the orbits of low mass circumstellar companions, we conduct combined analyses of the
radial velocity data as well as the Gaia and Hipparcos astrometric
data for eight nearby systems. Our study shows that companion-induced
position and proper motion differences between Gaia and Hipparcos are
significant enough to constrain orbits of low mass companions to a
precision comparable with previous combined analyses of direct imaging
and radial velocity data. We find that our method is robust to
  whether we use Gaia DR2 or Gaia EDR3, as well as whether we use all
  of the data, or just proper motion differences. In particular, we
fully characterize the orbits of HD 190360 b and HD 16160 C for the
first time. With a mass of 1.8$\pm$0.2\mj and an effective temperature of 123-176\,K and orbiting around a Sun-like
star, HD 190360 b is the smallest Jupiter-like planet with well-constrained mass and orbit, belonging to
a small sample of fully characterized Jupiter analogs. It is separated
from its primary star by 0.25$''$ and thus may be suitable for direct
imaging by the CGI instrument of the Roman Space Telescope. 
\end{abstract}
\section{Introduction}     \label{sec:introduction}
Since the discovery of the first objects accepted to be brown dwarfs:
GJ 229B \citep{nakajima95} and Teide 1 \cite{rebolo95}, hundreds of
brown dwarfs have been discovered. These objects are unlike main-sequence stars which are massive enough to
sustain nuclear fusion. \cite{saumon96} proposed the minimum mass
required to fuse deuterium (13\mj) as the lower mass limit of a brown
dwarf. Brown dwarfs are thought to form from the same process of molecular cloud
core collapse and fragmentation that forms stars, possibly resulting
in masses below 13\,$M_{\rm Jup}$ \citep{boss86,boss03}. Hence precise measurements of the mass of brown dwarfs are essential
to draw clearer lines between brown dwarfs, planets and stars in
order to study their population and formation scenarios. However,
rather than finding brown dwarf masses between the masses of stars and
planets, \cite{brandt20} found that the archetypal brown dwarf GJ 229
B as well as other T dwarfs have masses close to the minimum mass for
core hydrogen ignition.

Because most brown dwarfs are faint and typically on wide orbits
around stars, it takes long-term observations to constrain their orbits
and masses. Thanks to the progress with a variety of techniques, a few brown
dwarfs have been detected and characterized to a high precision through the radial
velocity (RV) technique \citep{patel07,quirrenbach19,rickman19},
multi-epoch direct imaging \citep{garcia17,dupuy19}, ground-based
astrometry \citep{dieterich18}, space-based astrometry
\citep{kervella19} and combinations thereof \citep{crepp16, brandt19a, grandjean19}. 

The two-decade baseline between Hipparcos
\citep{perryman97,leeuwen07} and Gaia astrometric surveys
\citep{gaia16b} means we are now in a better position to constrain the mass
and orbital elements of brown dwarfs on wide orbits. The offsets
between the proper motions measured by Gaia and Hipparcos have already
been used to constrain the mass of $\beta$ Pic \citep{snellen18}, HD
72946 B \citep{maire20}, and GJ 229 B \citep{brandt20}. In particular,
\cite{feng19c} detected $\epsilon$ Indi Ab through combined analysis
of the radial velocity observations and the difference in both
position and proper motion between the revised Hipparcos catalog
\citep{leeuwen07} and the Gaia Data Release 2 (DR2; \citep{gaia18}). This detection of a Jupiter
analog demonstrates the power of positional offsets in
constraining orbits of cold Jupiters.

In this work, we apply the combined approach introduced by
\cite{feng19c} in the analysis of radial velocity and Gaia-Hipparcos
astrometric offsets for eight nearby companions, including four M
dwarfs, three brown dwarfs, and one cold Jupiter. Based on the
consistency between our orbital solutions and previous results for the
stellar companions, we justify the use of our combined method in brown
dwarf mass and age estimation. In particular, we test the
  sensitivity of our approach to the choice of different versions of
  Gaia data and the effects of various systematics. This approach is
applied to estimate brown dwarf masses in order to understand their
evolution. The same method is used to determine the mass of a cold
Jupiter, providing a benchmark case for astrometric characterization of Jupiter analogs. 

The paper is structured as follows. We introduce the data in section
\ref{sec:data} and the methods in section \ref{sec:method}. The
results are presented and individual cases are discussed in section
\ref{sec:result}. We then study the evolution of three low mass
companions in section \ref{sec:BD} and conclude in section \ref{sec:conclusion}. 

\section{Data}\label{sec:data}
We collect the radial velocity (RV) data from various online archives
and publications including HARPS/ESO-3.6m \citep{pepe02}, CES/ESO
\citealt{kurster94}, SOPHIE/OHP \citep{perruchot11}, ELODIE/OHP
\citep{baranne96}, UVES/VLT \citep{dorn00}, HARPS-N/TNG, Keck/HIRES
\citep{vogt14}, Hamilton/Lick \citep{vogt87}, Levy/APF \citep{vogt94},
Whipple/AFOE \citep{brown94}, and spectrograph at the McDonald
Observatory (MCD; \citealt{cochran93}). In the case of archival HARPS-N and SOPHIE data we use TERRA \citep{anglada12} to process the
spectrum in order to get barycentric-corrected RVs. We use SERVAL to
reduce the HARPS data \citep{zechmeister18,trifonov20}. The
determination of long term RVs is a highly challenging activity which requires long term instrument stability. Most RV instruments
have data offsets present which are usually rather ill-characterised
and relatively poorly documented in the literature. For the case of
SOPHIE significant efforts have been made to correct
systematic instrumental drifts (e.g., \citealt{courcol15}) though are not
available as an offset for an arbitrary epoch. These corrections are anyways modest in comparison to the
errors on the SOPHIE measurements that we are using. In the case of
HARPS data there is a known offset in the RV zero point for the
post-2015 dataset \citep{curto15}. Thus we label the pre-2015
data set as ``HARPSpre'', and the post-2015 data as
``HARPSpost''. We also account for the changes of CCD detectors of
the Lick Hamilton spectrograph by using Lick6, Lick8, and Lick13 to
label the corresponding RV sets \citep{fischer14}. We show the sources and properties of all the RV data in Table \ref{tab:rv}. The
data for the new RVs used in this work will be available online. 
\begin{table}
    \centering
   \caption{Information for the data sets for the eight targets. The second column shows the data sets and the third column shows the corresponding source. In the two right columns, $N$ is the number of RVs per set and $\Delta T$ is the time span of a data set. For the new data used in this work, we
   cite the original instrumentation paper as the source. The new
     RV data used in this work will be available in the online verion.}
     \begin{tabular}{l l p{5cm} l r}
       \hline\hline
       Target& Instrument &Sources & N & $\Delta T$ [day]\\       \hline
       HD 131664 &HARPSpre&SERVAL&58&3713\\
             &HARPSpost&SERVAL&2&1\\\hline
       
HD 16160     &APF&\cite{vogt14} &159&1876\\
             &ELODIE/OHP &\url{http://atlas.obs-hp.fr/elodie/}&2&3545\\
             &HARPS-N&\url{http://archives.ia2.inaf.it/tng}&9&24\\
             &HARPSpre&SERVAL&230&1776\\       
             &HIRES/Keck & \cite{butler17}&116&6635\\
             &HAMILTON/Lick&\cite{fischer14}&62&8797\\\hline

HD 161797    & APF&\cite{vogt14}&21&1368\\
             & HIRES/Keck&\cite{butler17}&85&6262\\
             &HAMILTON/Lick&\cite{fischer14}&332&8797\\\hline

HD 190360    &AFOE&\cite{naef03}&13&1925\\
             &ELODIE/OHP&\url{http://atlas.obs-hp.fr/elodie/}&68&3940\\
             &HIRES/Keck&\cite{butler17}&235&6465\\
             &HAMILTON/Lick&\cite{fischer14}&149&3334\\
             &SOPHIE/OHP&\url{http://atlas.obs-hp.fr/sophie/} archive data reprocessed with TERRA&30&469\\\hline

HD 190406 &APF&\cite{vogt14}&83&2215\\
             &Keck&\cite{butler17}&203&7497\\
              &HAMILTON/Lick&\cite{fischer14}&125&8799\\
       
HD 39587     &CFHT&\cite{walker95}&38&3621\\
             &HAMILTON/Lick&\cite{fischer14}&38&4579\\
             &SOPHIE/OHP&\url{http://atlas.obs-hp.fr/sophie/} archive data reprocessed with TERRA&637&728\\\hline

HD 4747      & C07 &CORALIE \citep{quiloz00} after first update in 2007&7&449\\
             & C98 &CORALIE \citep{quiloz00} before first update in 2007&29&2740\\
             &HIRES/Keck&This work&50&6546\\\hline
       
HIP 2552     &ELODIE/OHP&\url{http://atlas.obs-hp.fr/elodie/}&20&1921\\\hline

     \end{tabular}
     \label{tab:rv}
\end{table}

For the sample of targets shown in Table \ref{tab:rv}, we collect
astrometry data from the revised Hipparcos catalog \citep{leeuwen07}. we use the {\small gaiadr2.tmass\_best\_neighbour}
cross-matching catalog in the Gaia data archive to find the
designation of Gaia DR2 and use the
{\small gaiaedr3.dr2\_neighbourhood} cross-matching catalog to find
the data from Gaia Early Data Release 3 (EDR3; \citealt{gaia20}). For a target without Gaia counterparts in the
cross-matching catalogs, we select the Gaia sources within 0.1 degree from
its Hipparcos coordinates and with a parallax differing
from the Hipparcos one by less than 10\%. For stars with both DR2 and
EDR3 data, we use the difference between the revised Hipparcos catalog and the Gaia EDR3 to
constrain the orbits of companions. We test the sensitivity of
the orbital solutions to the choice of different Gaia catalogs in
section \ref{sec:sensitivity}. 

\begin{table}
  \caption{
Hipparcos and Gaia EDR3 catalog astrometry for the sample of
stars. The mass of HD 161797 is from \protect{\citep{grundahl17}}
while the masses of other stars are from the TESS input catalog
\citep{stassun19}. The superscripts ``gaia'' and ``hip'' are
respectively used to denote the Gaia EDR3 and Hipparcos right
 ascension ($\alpha$), declination ($\delta$), parallax ($\varpi$), proper motion
  in right ascension ($\mu_\alpha$), and the proper motion in
  declination ($\mu_\delta$).}
  \begin{tabular}{llll llll lllr}
    \hline\hline
Star&Mass&$\alpha^{\rm hip}$&$\delta^{\rm hip}$&$\varpi^{\rm
                                                hip}$&$\mu_\alpha^{\rm
                                                        hip}$&$\mu_\delta^{\rm
                                                               hip}$&$\alpha^{\rm gaia}$&$\delta^{\rm gaia}$&$\varpi^{\rm gaia}$&$\mu_\alpha^{\rm gaia}$&$\mu_\delta^{\rm gaia}$ \\ 
&$M_\odot$&deg&deg&mas&\masyr&\masyr&deg&deg&mas&\masyr&\masyr\\
\hline
HD 131664 & 1.1(1) &225.03&-73.5&17.8(7)&14.6(6)&28.8(6)&225.03&-73.54&19.14(8)&8.01(8)&22.28(9) \\ 
HD 16160 & 0.78(9) &39.02&6.9&139.3(4)&1807.8(9)&1444.0(4)&39.0&6.9&138.3(3)&1778.6(4)&1477.3(2) \\ 
HD 161797 & 1.11(1) &266.615&27.7&120.3(2)&-291.7(1)&-749.6(2)&266.6&27.7&119.9(2)&-312.1(1)&-773.2(2) \\ 
HD 190360 & 1.0(1) &300.90&29.9&63.1(3)&683.9(2)&-524.7(3)&300.91&29.89&62.49(4)&683.20(3)&-525.50(4) \\ 
HD 190406 & 1.1(1) &301.03&17.1&56.3(4)&-394.6(3)&-407.8(3)&301.02&17.07&56.27(4)&-387.47(4)&-419.50(3) \\ 
HD 39587 & 1.1(1) &88.60&20.3&115.4(3)&-162.5(3)&-99.5(2)&88.6&20.3&114.9(5)&-179.0(4)&-90.7(3) \\ 
HD 4747 & 0.9(1) &12.36&-23.2&53.5(5)&516.9(6)&120.0(4)&12.36&-23.21&53.05(3)&519.05(3)&124.04(3) \\ 
HIP 2552 & 0.53(8) &8.1&67&99(2)&1738(2)&-224(2)&8.1&67.2&101.1(5)&1748.3(4)&-347.6(5) \\ 
    \hline
  \end{tabular}
  \label{tab:astrometry}
\end{table}

\section{Numerical and statistical methods}\label{sec:method}
Considering that the combined model of astrometry and RV was introduced by
\cite{feng19c}, we briefly introduce the model and method here. The Keplerian part of the RV model is
\begin{equation}
  v^{\rm kep}=K \left[\cos{(\omega_\star+\nu)}+e\cos{\omega_\star}\right]+b~,
  \label{eqn:vr}
\end{equation}
where $K$ is the semi-amplitude of the Keplerian RV variation,
$\omega_\star$ is the argument of periastron of the stellar reflex motion,
$\nu$ is the true anomaly, $e$ is the eccentricity of the planetary
orbit, and $b$ is the RV offset. The semi-amplitude is
\begin{equation}
  K=v\sin{I}=\sqrt{\dfrac{Gm_p^2}{m_p+m_\star}\dfrac{1}{a(1-e^2)}}\sin{I}~,
  \label{eqn:K}
\end{equation}
where $v$ is the companion-induced velocity, $a$ is the semi-major
axis of the planet's relative orbit, $I$ is the inclination, and $m_p$ and
$m_\star$ are the masses of planet and star,
respectively\footnote{Note that equation \ref{eqn:K} in this paper is
  correct while the square root for
  $\dfrac{Gm_p^2}{m_p+m_\star}\dfrac{1}{a(1-e^2)}$ in equation 5 in
  \cite{feng19c} is missing}.

The astrometry model of position and proper motion $\bm{\hat\eta}\equiv
[\hat\alpha,\hat\delta,\hat\mu_\alpha,\hat\mu_\delta]$ at the Gaia and
Hipparcos reference epochs is already introduced by \cite{feng19c},
and is thus not repeated here. The likelihood of the combined RV and astrometry model is
\begin{align}
  \mathcal{L}&\equiv P(D|\theta,M)=\prod_k^{N_{\rm
      set}}\prod_i^{N_k^{\rm
      rv}}\frac{1}{\sqrt{2\pi(\sigma_j^2+\sigma_{\rm
        Jk}^2)}}\exp\left\{-\frac{[v_r(t_i)-\hat{v}_r(t_i)]^2}{2(\sigma_i^2+\sigma_{\rm Jk}^2)}\right\}\nonumber\\
  &+(2\pi)^{\frac{N_{\rm epoch}N_{\rm par}}{2}}\prod_j^{N_{\rm epoch}}({\rm det}\Sigma_j)^{-\frac{1}{2}}{\rm exp}\left\{-\frac{1}{2}[\hat{\bm{\eta}}(t_i)-\bm{\eta}(t_i)]^T\Sigma_j^{-1}[\hat{\bm{\eta}}(t_i)-\bm{\eta}(t_i)]\right\}~,
\label{eqn:comblike}
\end{align}
where $N_{\rm set}$, $N_{\rm epoch}$, and $N_{\rm par}$  are
respectively the number of RV data sets, astrometry epochs, and free
parameters of the astrometry model. $N_k^{\rm rv}$ is the number of
RVs in the $k^{\rm th}$ RV set, $\bm{\eta}\equiv
[\alpha,\delta,\mu_\alpha,\mu_\delta]$ is the astrometry data, and
$\Sigma$ is the covariance matrix of $\bm{\eta}$ corrected by jitter,
$\Sigma_j\equiv \Sigma_{\rm 0j}(1+J_j)$ where $\Sigma_{\rm 0j}$ is the
catalog covariance matrix for the $j^{\rm th}$ astrometry epoch, and
$J_j$ is the so-called ``relative astrometry jitter''. The order of
the MA
model is chosen in the Bayesian framework such that the order is
chosen to avoid false negatives and false positives
\citep{feng17a}. We adopt uniform priors for $\ln{P}$, $\ln{\tau}$,
and the other parameters.

We only consider one-companion models in this work
because we choose targets with only one dominant RV and astrometry signal. Hence the
free parameters in the combined RV and astrometry model are orbital period ($P$), RV semi-smplitude ($K$), eccentricity
($e$), inclination ($I$), argument of periastron ($\omega_\star$),
longitude of ascending node ($\Omega$), and mean anomaly ($M_0$) at the
reference epoch $t_0$ which is the earliest epoch of the RV data, RV
jitter ($\sigma_J$), time scale ($\tau$) and
amplitude ($w$) of the MA model, offset in $\alpha$
($\Delta\alpha$), offset in $\delta$
($\Delta\delta$), offset in $\mu_\alpha$ ($\Delta\mu_\alpha$), offset in
$\mu_\delta$ ($\Delta\mu_\delta$), logarithmic jitter in Gaia astrometry
($\ln{J_{\rm gaia}}$), and logarithmic jitter in Hipparcos astrometry
($\ln{J_{\rm hip}}$). The companion mass ($m_p$), semi-major axis ($a$),
and the epoch at the periastron ($T_P$) are derived from the free
parameters by adopting the stellar masses in Table
\ref{tab:astrometry}. 

{\ We use four offset parameters, $\Delta\alpha$, $\Delta\delta$,
 $\Delta\mu_\alpha$, and $\Delta\mu_\delta$, and astrometry
 jitters, $J_{\rm gaia}$ and $J_{\rm hip}$, to account for the frame
 rotation between Gaia and Hipparcos catalogs
 \citep{lindegren18,brandt18}, zero point offsets in parallaxes
 \citep{lindegren18}, proper motion offsets caused by light travel
 time \citep{kervella19}, and other effects. One type of
   systematics is caused by the deviation of catalog position and
   proper motion from instantaneous values at the Gaia and Hipparcos
   reference epochs. Because the motion of a star induced by its companion
   is nonlinear, the fitting of a five-parameter astrometry model to
   the astrometry data for this star leads to a bias. This bias is
   caused by model incompleteness and is approximately the offsets
   caused by a companion at the central epoch of the data
   baseline. For a system with its barycenter static to the observer,
   the astrometry signal of the primary star caused by a wide
   companion could be explained by a fit of a linear function to a
   well sampled arc. The slope of the line is equivalent to the proper
   motion of the star at the central epoch of the data baseline. The
   center of the line measures the position of the star at the central
   epoch, which deviates from the instantaneous position (equivalent
   to the center of the arc). We call this the ``smearing effect''. For a circular orbit on the sky, this
   positional bias is about $(\pi\Delta t/P)^2/2$, which is about 10\%
of the semi-major axis for a 20-year (i.e. $P=20$\,yr) orbit observed
over $\Delta t=$3\,yr. Considering a combined fit to positional difference together with
    proper motion difference and RV data, this smearing effect would be
   significantly reduced. Although the GOST tool\footnote{\url{https://gaia.esac.esa.int/gost/}} is able to generate
   synthetic Gaia data in order to minimize the smearing effect, it
   could be time consuming when combined with posterior sampling, and
   an investigation of its feasibility is beyond the scope of this work. 

Moreover, \cite{brandt18} finds a 0.1-0.2 year difference between the central and
reference epochs for the Gaia and Hipparcos catalogs. Considering that
the proper motion and positional biases due to this timing difference
and the smearing effect are indistinguishable
from those caused by other effects, we use the four offset parameters
and two astrometry jitters to account for all of these effects {\it a posteriori}. While
previous studies investigate these effects separately and globally (e.g.,
\citealt{kervella19} and \citealt{brandt19a}), we model them and the companion signal
 simultaneously for individual stars in order to avoid over-fitting or
 under-fitting problems caused by separating calibration from signal
 search (e.g., \citealt{foreman-mackey15}). To further
   justify our methodology, we will compare our method
  with other methods and test the sensitivity of our orbital solutions
  to various systematics in section \ref{sec:sensitivity}.

To explore the posterior of the full
model, we use adaptive Markov Chain Monte Carlo (MCMC) in combination with
parallel computation techniques \citep{haario01,feng19a}. Furthermore, we launch multiple
tempered chains to explore the global posterior distribution in order
to find global posterior maximum. Then we launch non-tempered chains to
explore the global maximum. We use the Gelman-Rubin criterion to
measure the convergence of a chain \citep{gelman14}. To measure the
significance of a signal, we calculate the Bayes factor (BF) using the
method introduced by \cite{feng16} and select signals using
$\ln{\rm BF}>5$.

\section{Combined analyses of radial velocity and astrometric data}\label{sec:result}
We apply the combined model to eight nearby stars that have high
precision RVs as well as astrometry from Hipparcos and Gaia EDR3. Since
these systems have known planets or companions, the RV signal is very
significant and thus we use the white noise model with offsets to
model the RV data set. The masses and orbital parameters of the
  eight low mass companions based on the combined RV and astrometry
  analyses are shown in Table \ref{tab:edr3}. In Fig. \ref{fig:fit1} and Fig. \ref{fig:fit2},
we show the best orbital solutions of combined modeling for the eight
stellar systems.
\begin{landscape}
\begin{table}
\centering
\caption{Orbital parameters for stellar and planetary companions based on combined RV and astrometric analyses. Among the parameters, directly inferred parameters are $P$ (orbital period), $K$ (RV semi-amplitude), $e$ (eccentricity), $\omega_\star$ (argument of periastron), $M_0$ (mean anomaly at the minimum epoch of RV data), $I$ (inclination), $\Omega$ (longitude of ascending node), ln$J_{\rm hip}$ (logarithmic jitter in Hipparcos astrometry), ln$J_{\rm gaia}$ (logarithmic jitter in Gaia astrometry), $\Delta \alpha$ (right ascension offset), $\Delta \delta$ (declination offset), $\Delta \mu_\alpha$ (proper motion in right ascension offset), and $\Delta \mu_\delta$ (proper motion in declination offset). The derived parameters are T$_p$ (periastron epoch), $m_p$ (planetary mass), and $a$ (semi-major axis). For each parameter, the upper row shows the mean and standard deviation of the posterior samples, and the lower row shows the {\it maximum a posteriori} (MAP) value with the 1\% and 99\% quantiles.  }
\begin{tabular}{lrrrr rrrr}
\hline
\hline
Name & HD 131664 B & HD 16160 C & HD 161797 Ab & HD 190360 b & HD 190406 B & HD 39587 B & HD 4747 B & HIP 2552 C \\
Other Name & HIP 73408 B & GJ 105 C & $\mu$ Herculis Ab & GJ 777 b & GJ 779 B & $\chi^1$ Orionis B & GJ 36 B & GJ 22 C \\\hline
$P$ [yr] & $ 5.424 \pm 0.004 $ & $ 76.107 \pm 1.820 $ & $ 81.610 \pm 8.849 $ & $ 7.815 \pm 0.035 $ & $ 61.674 \pm 1.554 $ & $ 14.115 \pm 0.005 $ & $ 33.229 \pm 0.576 $ & $ 15.558 \pm 0.097 $ \\
 & $ 5.426 _{- 0.012 }^{+ 0.007 }$ & $ 77.746 _{- 5.439 }^{+ 2.776 }$ & $ 94.349 _{- 32.403 }^{+ 1.072 }$ & $ 7.841 _{- 0.107 }^{+ 0.057 }$ & $ 63.071 _{- 4.778 }^{+ 2.725 }$ & $ 14.116 _{- 0.009 }^{+ 0.012 }$ & $ 33.179 _{- 1.225 }^{+ 1.516 }$ & $ 15.574 _{- 0.276 }^{+ 0.161 }$ \\
$K$ [m/s] & $ 433.7 \pm 3.1 $ & $ 711.7 \pm 2.2 $ & $ 1183.7 \pm 73.2 $ & $ 24.4 \pm 0.6 $ & $ 537.7 \pm 2.9 $ & $ 1888.3 \pm 4.5 $ & $ 703.2 \pm 10.0 $ & $ 2219.0 \pm 12.7 $ \\
 & $ 434.2 _{- 7.0 }^{+ 6.2 }$ & $ 710.5 _{- 3.9 }^{+ 6.3 }$ & $ 1275.3 _{- 266.4 }^{+ 1.6 }$ & $ 24.2 _{- 1.0 }^{+ 1.6 }$ & $ 539.1 _{- 8.6 }^{+ 5.4 }$ & $ 1883.9 _{- 5.0 }^{+ 14.9 }$ & $ 704.7 _{- 23.8 }^{+ 22.9 }$ & $ 2234.8 _{- 49.7 }^{+ 8.5 }$ \\
$e$ & $ 0.693 \pm 0.002 $ & $ 0.641 \pm 0.004 $ & $ 0.386 \pm 0.025 $ & $ 0.340 \pm 0.018 $ & $ 0.462 \pm 0.007 $ & $ 0.450 \pm 0.002 $ & $ 0.732 \pm 0.002 $ & $ 0.134 \pm 0.015 $ \\
 & $ 0.692 _{- 0.005 }^{+ 0.006 }$ & $ 0.645 _{- 0.014 }^{+ 0.007 }$ & $ 0.418 _{- 0.089 }^{+ 0.002 }$ & $ 0.338 _{- 0.042 }^{+ 0.042 }$ & $ 0.468 _{- 0.020 }^{+ 0.011 }$ & $ 0.451 _{- 0.005 }^{+ 0.003 }$ & $ 0.731 _{- 0.004 }^{+ 0.005 }$ & $ 0.142 _{- 0.039 }^{+ 0.028 }$ \\
$\omega_\star$ [deg] & $ 151.8 \pm 0.3 $ & $ 133.4 \pm 0.6 $ & $ 228.3 \pm 2.4 $ & $ 19.9 \pm 3.6 $ & $ 267.6 \pm 1.1 $ & $ 110.6 \pm 0.4 $ & $ 266.8 \pm 0.5 $ & $ 275.5 \pm 6.1 $ \\
 & $ 151.6 _{- 0.5 }^{+ 0.8 }$ & $ 132.9 _{- 0.8 }^{+ 1.9 }$ & $ 225.6 _{- 1.2 }^{+ 8.6 }$ & $ 20.2 _{- 8.5 }^{+ 7.8 }$ & $ 266.6 _{- 1.6 }^{+ 3.8 }$ & $ 110.2 _{- 0.9 }^{+ 1.1 }$ & $ 266.6 _{- 1.0 }^{+ 1.4 }$ & $ 265.4 _{- 0.4 }^{+ 24.3 }$ \\
$M_0$ [deg] & $ 152.9 \pm 0.3 $ & $ 215.3 \pm 3.5 $ & $ 200.6 \pm 18.6 $ & $ 219.5 \pm 4.1 $ & $ 199.3 \pm 4.3 $ & $ 265.9 \pm 0.4 $ & $ 356.9 \pm 0.2 $ & $ 8.3 \pm 5.5 $ \\
 & $ 153.2 _{- 1.0 }^{+ 0.6 }$ & $ 218.5 _{- 10.8 }^{+ 4.9 }$ & $ 223.7 _{- 70.1 }^{+ 1.5 }$ & $ 220.7 _{- 10.8 }^{+ 8.3 }$ & $ 203.3 _{- 14.0 }^{+ 6.8 }$ & $ 266.0 _{- 0.8 }^{+ 0.8 }$ & $ 356.9 _{- 0.5 }^{+ 0.4 }$ & $ 17.6 _{- 22.0 }^{+ 0.1 }$ \\
$I$ [deg] & $ 170.7 \pm 1.5 $ & $ 45.5 \pm 2.8 $ & $ 58.8 \pm 2.4 $ & $ 80.2 \pm 23.2 $ & $ 90.7 \pm 5.0 $ & $ 93.4 \pm 1.9 $ & $ 50.6 \pm 2.0 $ & $ 47.9 \pm 0.9 $ \\
 & $ 171.3 _{- 8.0 }^{+ 1.1 }$ & $ 47.6 _{- 7.9 }^{+ 3.2 }$ & $ 61.8 _{- 8.7 }^{+ 0.3 }$ & $ 65.0 _{- 22.3 }^{+ 66.3 }$ & $ 94.1 _{- 15.0 }^{+ 5.9 }$ & $ 93.5 _{- 6.0 }^{+ 3.9 }$ & $ 51.0 _{- 6.5 }^{+ 4.5 }$ & $ 48.6 _{- 2.4 }^{+ 1.9 }$ \\
$\Omega$ [deg] & $ 348.3 \pm 6.4 $ & $ 122.5 \pm 3.4 $ & $ 270.4 \pm 4.2 $ & $ 306.6 \pm 44.4 $ & $ 146.5 \pm 3.3 $ & $ 128.0 \pm 5.5 $ & $ 273.0 \pm 3.8 $ & $ 177.8 \pm 1.6 $ \\
 & $ 347.4 _{- 10.5 }^{+ 21.5 }$ & $ 125.6 _{- 8.7 }^{+ 3.6 }$ & $ 265.2 _{- 0.6 }^{+ 15.3 }$ & $ 290.0 _{- 89.7 }^{+ 122.6 }$ & $ 148.4 _{- 9.3 }^{+ 4.4 }$ & $ 127.8 _{- 8.8 }^{+ 22.7 }$ & $ 272.1 _{- 6.0 }^{+ 16.3 }$ & $ 177.8 _{- 4.9 }^{+ 3.0 }$ \\
$m_p$ [$M_{\rm Jup}$] & $ 127.8 \pm 17.9 $ & $ 102.6 \pm 9.4 $ & $ 231.9 \pm 16.8 $ & $ 1.8 \pm 0.2 $ & $ 72.8 \pm 6.1 $ & $ 167.2 \pm 13.0 $ & $ 68.3 \pm 5.9 $ & $ 198.9 \pm 19.5 $ \\
 & $ 135.3 _{- 65.6 }^{+ 28.1 }$ & $ 99.1 _{- 17.4 }^{+ 26.5 }$ & $ 253.9 _{- 62.2 }^{+ 1.9 }$ & $ 1.7 _{- 0.4 }^{+ 0.7 }$ & $ 73.4 _{- 15.3 }^{+ 13.1 }$ & $ 167.0 _{- 31.0 }^{+ 29.4 }$ & $ 68.1 _{- 13.7 }^{+ 13.9 }$ & $ 199.0 _{- 47.3 }^{+ 43.1 }$ \\
$a$ [au] & $ 3.2 \pm 0.1 $ & $ 17.0 \pm 0.7 $ & $ 20.5 \pm 1.6 $ & $ 3.9 \pm 0.2 $ & $ 16.3 \pm 0.7 $ & $ 6.3 \pm 0.2 $ & $ 10.2 \pm 0.4 $ & $ 5.4 \pm 0.3 $ \\
 & $ 3.2 _{- 0.3 }^{+ 0.3 }$ & $ 17.3 _{- 2.0 }^{+ 1.3 }$ & $ 22.8 _{- 5.9 }^{+ 0.1 }$ & $ 3.9 _{- 0.4 }^{+ 0.3 }$ & $ 16.6 _{- 2.1 }^{+ 1.4 }$ & $ 6.3 _{- 0.6 }^{+ 0.5 }$ & $ 10.2 _{- 1.1 }^{+ 0.9 }$ & $ 5.5 _{- 0.7 }^{+ 0.5 }$ \\
$T_p$-2400000 [day] & $ 52023 \pm 2 $ & $ 30414 \pm 669 $ & $ 30277 \pm 3218 $ & $ 47871 \pm 38 $ & $ 34570 \pm 586 $ & $ 41150 \pm 7 $ & $ 38335 \pm 213 $ & $ 51030 \pm 1696 $ \\
 & $ 52021 _{- 4 }^{+ 7 }$ & $ 29810 _{- 1016 }^{+ 2001 }$ & $ 25631 _{- 389 }^{+ 11748 }$ & $ 47856 _{- 75 }^{+ 102 }$ & $ 34036 _{- 1019 }^{+ 1811 }$ & $ 41148 _{- 14 }^{+ 15 }$ & $ 38352 _{- 558 }^{+ 458 }$ & $ 51496 _{- 5417 }^{+ 276 }$ \\
ln$J_{\rm gaia}$ & $ -0.28 \pm 3.87 $ & $ 2.29 \pm 3.16 $ & $ -3.24 \pm 4.32 $ & $ -3.84 \pm 4.17 $ & $ -3.85 \pm 4.20 $ & $ 0.41 \pm 3.73 $ & $ -1.97 \pm 5.09 $ & $ -0.96 \pm 0.26 $ \\
 & $ -0.71 _{- 7.86 }^{+ 7.00 }$ & $ -5.30 _{- 1.59 }^{+ 11.80 }$ & $ -6.96 _{- 4.24 }^{+ 11.16 }$ & $ -5.39 _{- 5.90 }^{+ 9.94 }$ & $ -8.42 _{- 2.83 }^{+ 13.04 }$ & $ -8.22 _{- 1.86 }^{+ 14.18 }$ & $ -6.33 _{- 4.79 }^{+ 13.77 }$ & $ -1.04 _{- 0.39 }^{+ 0.71 }$ \\
ln$J_{\rm hip}$ & $ 3.99 \pm 1.15 $ & $ -1.40 \pm 4.49 $ & $ 3.75 \pm 1.88 $ & $ 1.97 \pm 1.85 $ & $ 2.40 \pm 1.29 $ & $ 3.48 \pm 2.25 $ & $ -3.13 \pm 4.10 $ & $ -3.92 \pm 0.27 $ \\
 & $ 4.40 _{- 2.65 }^{+ 3.02 }$ & $ 0.46 _{- 11.60 }^{+ 5.30 }$ & $ -0.05 _{- 0.00 }^{+ 8.06 }$ & $ 0.67 _{- 6.93 }^{+ 5.00 }$ & $ 1.33 _{- 1.87 }^{+ 4.47 }$ & $ 4.09 _{- 5.88 }^{+ 4.98 }$ & $ -1.87 _{- 9.36 }^{+ 5.59 }$ & $ -4.18 _{- 0.31 }^{+ 0.83 }$ \\
$\Delta\alpha$ [mas] & $ -0.03 \pm 0.26 $ & $ 0.33 \pm 1.95 $ & $ -0.05 \pm 0.19 $ & $ 0.00 \pm 0.05 $ & $ 0.00 \pm 0.04 $ & $ -0.09 \pm 1.32 $ & $ -0.01 \pm 0.18 $ & $ 0.09 \pm 0.18 $ \\
 & $ 0.01 _{- 0.87 }^{+ 0.66 }$ & $ -0.10 _{- 4.80 }^{+ 6.00 }$ & $ 0.03 _{- 0.70 }^{+ 0.33 }$ & $ 0.02 _{- 0.26 }^{+ 0.07 }$ & $ 0.00 _{- 0.14 }^{+ 0.09 }$ & $ -0.44 _{- 4.01 }^{+ 3.76 }$ & $ -0.01 _{- 0.60 }^{+ 0.57 }$ & $ -0.11 _{- 0.20 }^{+ 0.59 }$ \\
$\Delta\delta$ [mas] & $ -0.07 \pm 0.30 $ & $ -0.10 \pm 1.36 $ & $ 0.04 \pm 0.28 $ & $ -0.01 \pm 0.05 $ & $ 0.00 \pm 0.04 $ & $ -0.19 \pm 1.38 $ & $ 0.02 \pm 0.13 $ & $ 0.37 \pm 0.37 $ \\
 & $ -0.01 _{- 1.23 }^{+ 0.54 }$ & $ 0.21 _{- 4.54 }^{+ 3.43 }$ & $ -0.06 _{- 0.68 }^{+ 0.95 }$ & $ 0.01 _{- 0.28 }^{+ 0.09 }$ & $ 0.00 _{- 0.10 }^{+ 0.14 }$ & $ -0.24 _{- 3.24 }^{+ 5.15 }$ & $ 0.00 _{- 0.25 }^{+ 0.70 }$ & $ 0.67 _{- 1.10 }^{+ 0.62 }$ \\
$\Delta\mu_\alpha$ [mas/yr] & $ 0.00 \pm 0.36 $ & $ 2.56 \pm 2.00 $ & $ -0.06 \pm 0.28 $ & $ 0.00 \pm 0.04 $ & $ 0.00 \pm 0.07 $ & $ -0.92 \pm 0.81 $ & $ 0.01 \pm 0.14 $ & $ 0.02 \pm 0.20 $ \\
 & $ 0.04 _{- 1.48 }^{+ 0.94 }$ & $ 0.68 _{- 1.80 }^{+ 5.36 }$ & $ -0.06 _{- 1.24 }^{+ 0.50 }$ & $ 0.01 _{- 0.10 }^{+ 0.09 }$ & $ -0.03 _{- 0.13 }^{+ 0.28 }$ & $ -0.77 _{- 2.75 }^{+ 1.44 }$ & $ 0.00 _{- 0.35 }^{+ 0.53 }$ & $ -0.01 _{- 0.37 }^{+ 0.46 }$ \\
$\Delta\mu_\delta$ [mas/yr] & $ -0.30 \pm 0.54 $ & $ -0.61 \pm 1.33 $ & $ -0.07 \pm 0.23 $ & $ -0.02 \pm 0.05 $ & $ 0.01 \pm 0.06 $ & $ 0.46 \pm 0.93 $ & $ 0.07 \pm 0.23 $ & $ -0.04 \pm 0.56 $ \\
 & $ -0.06 _{- 2.19 }^{+ 0.38 }$ & $ 0.00 _{- 4.10 }^{+ 2.06 }$ & $ -0.01 _{- 0.82 }^{+ 0.47 }$ & $ -0.02 _{- 0.18 }^{+ 0.09 }$ & $ 0.00 _{- 0.10 }^{+ 0.26 }$ & $ 0.33 _{- 2.81 }^{+ 2.28 }$ & $ 0.01 _{- 0.24 }^{+ 1.11 }$ & $ 0.15 _{- 1.65 }^{+ 0.97 }$ \\
\hline
\end{tabular}
\label{tab:edr3}
\end{table}
\end{landscape}
\begin{figure}
\includegraphics[scale=0.6]{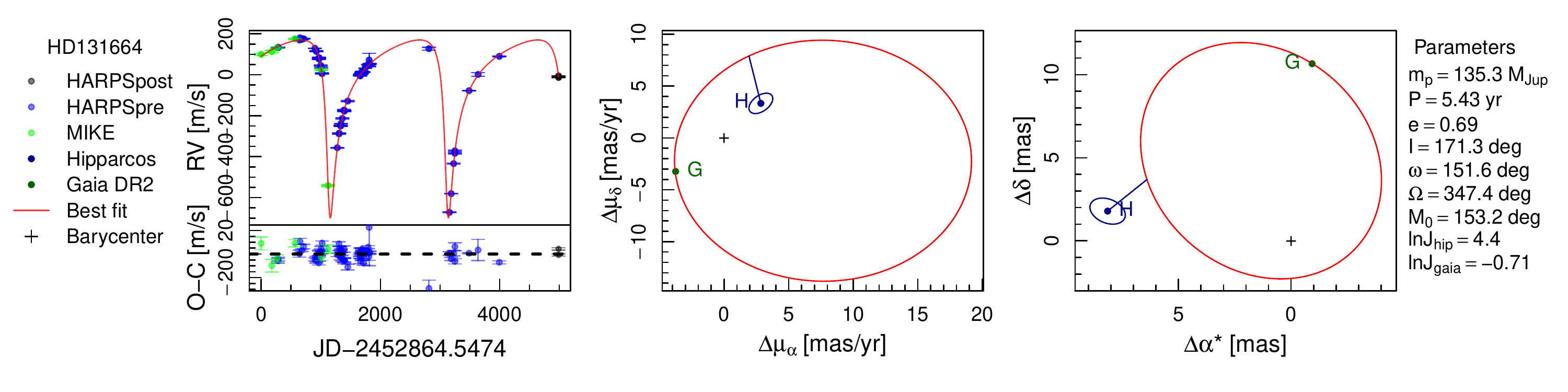}\vspace*{-0.2in}
\includegraphics[scale=0.6]{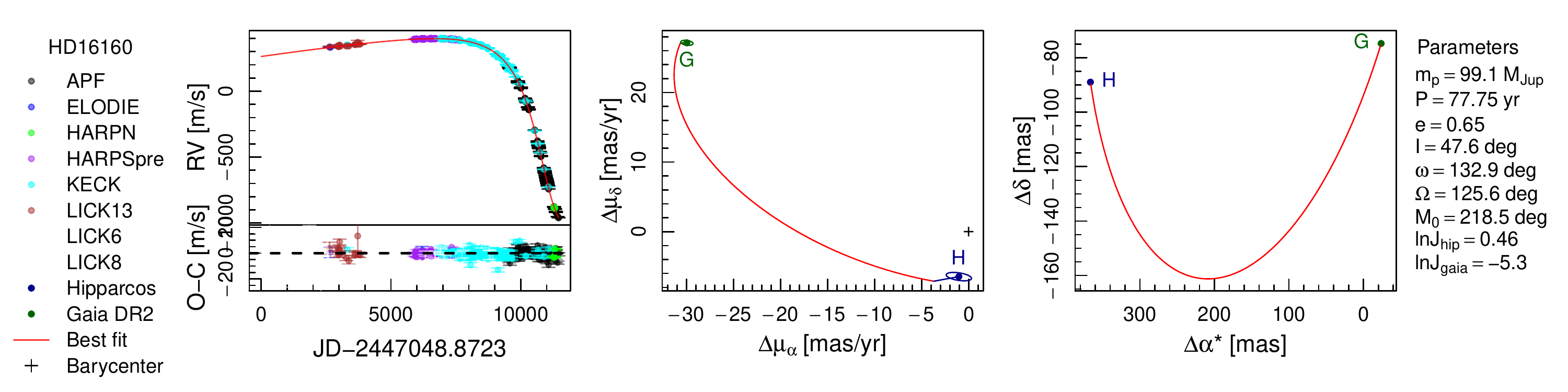}\vspace*{-0.2in}
\includegraphics[scale=0.6]{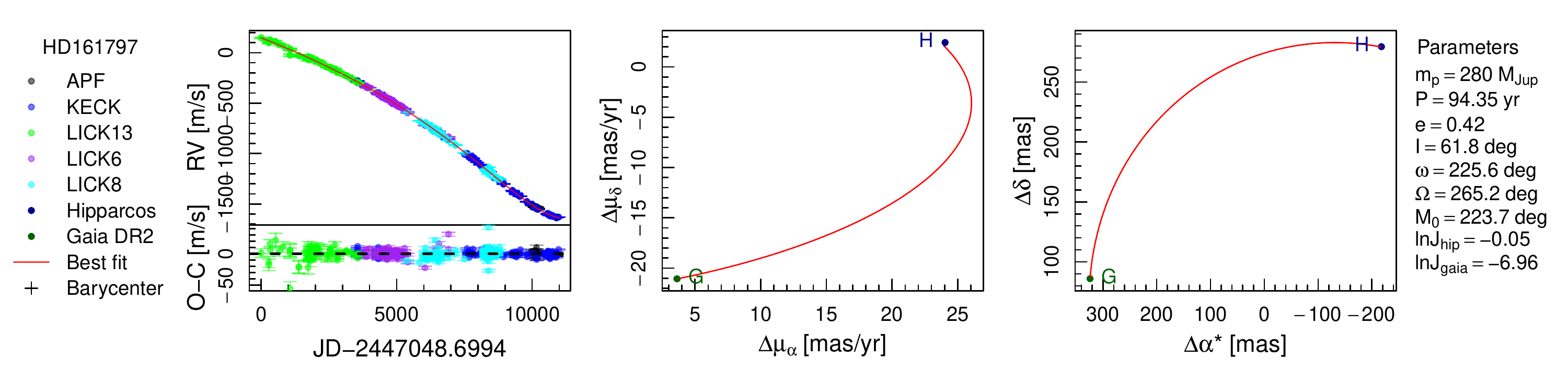}\vspace*{-0.2in}
\includegraphics[scale=0.6]{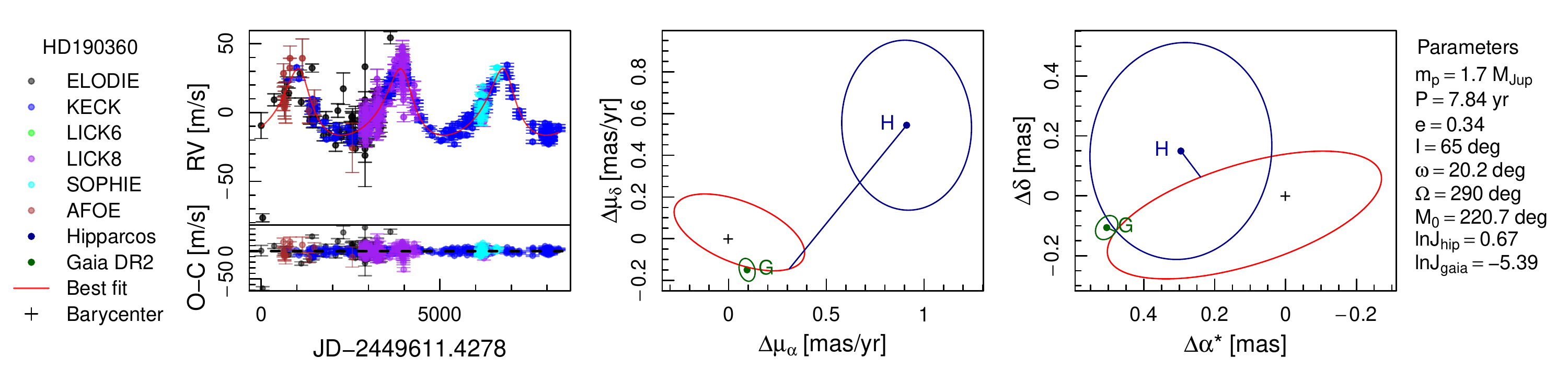}\vspace*{-0.2in}
\caption{Combined RV and astrometry fit for HD 16160, HD 161797, HD182488, and HD 190360. The left panels show the RV fits, the middle ones show proper motion fit, and the right ones show position fit. For each target, the linear motion is subtracted from the proper motion and position to show planet-induced nonlinear motion. The RV data sets as well as Gaia and Hipparcos astrometric data are encoded by the same colors as shown by the left-hand legends. The covariances of the Gaia and Hipparcos proper motions and positions are denoted by error ellipses. The straight line connects the data point to the best fitting model, which represents where the companion is expected to be. The parameters at the MAP are shown on the right side of the figure. }
\label{fig:fit1}
\end{figure}

\begin{figure}
\includegraphics[scale=0.6]{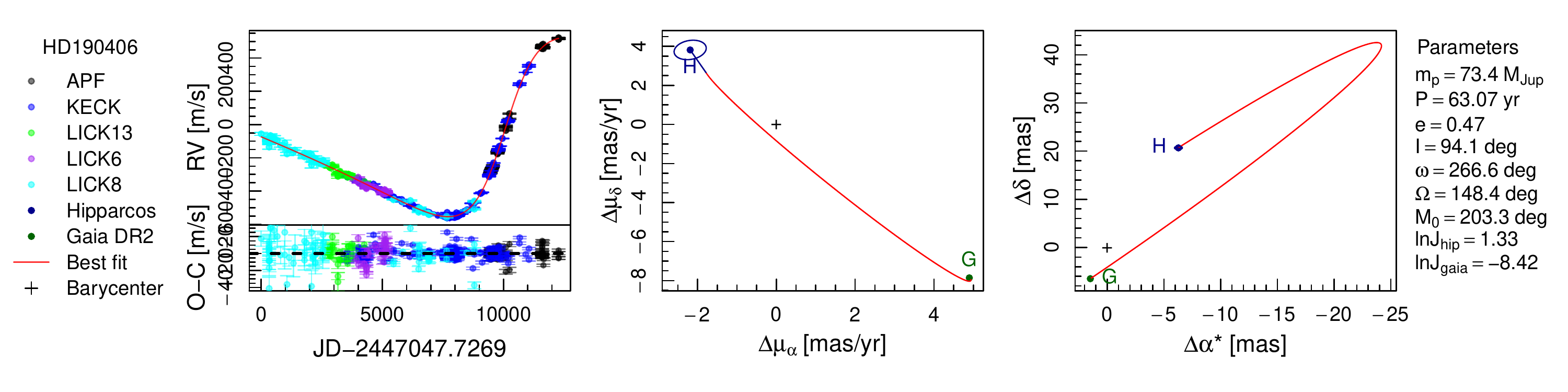}\vspace*{-0.2in}
\includegraphics[scale=0.6]{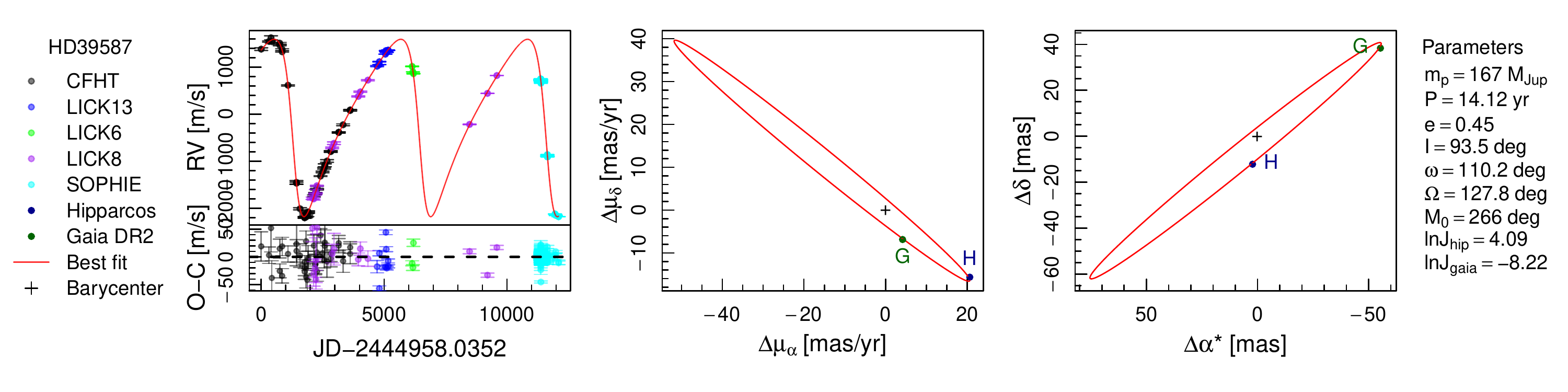}\vspace*{-0.2in}
\includegraphics[scale=0.6]{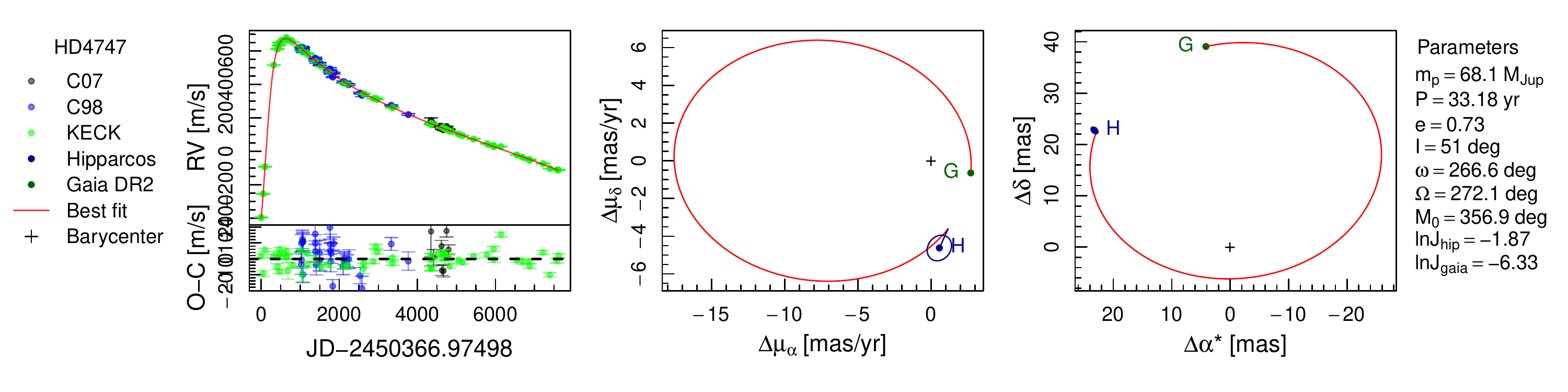}\vspace*{-0.2in}
\includegraphics[scale=0.6]{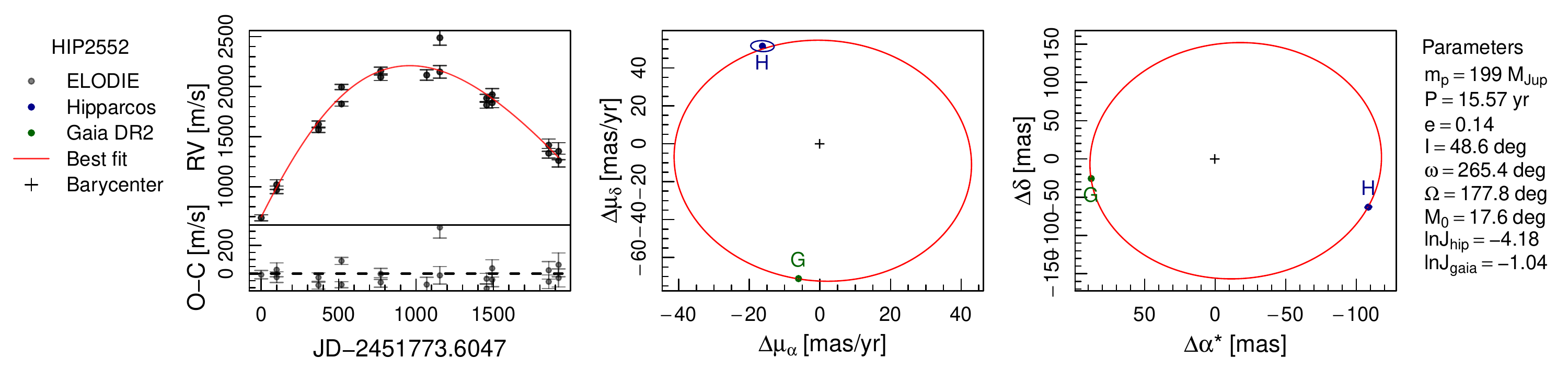}
\caption{Similar to Fig. \ref{fig:fit1} but for other targets.}
\label{fig:fit2}
\end{figure}

As seen from Fig. \ref{fig:comparison}, our solutions for HD 131664,
HD 190406 B, HD 4747 B, HIP 2552 C, and HD 39587 B are consistent with previous
solutions. Since we include stellar mass uncertainty in our estimates,
our estimates of the companion masses are more conservative
than many previous efforts. Thanks to the combined analysis and
in particular the astrometry data, we find full orbital solutions and
dynamical masses for HD 190360 b and HD 16160 C, compared
with partial or no orbital solutions in the literature.

\begin{figure}
  \centering
  \includegraphics[scale=0.6]{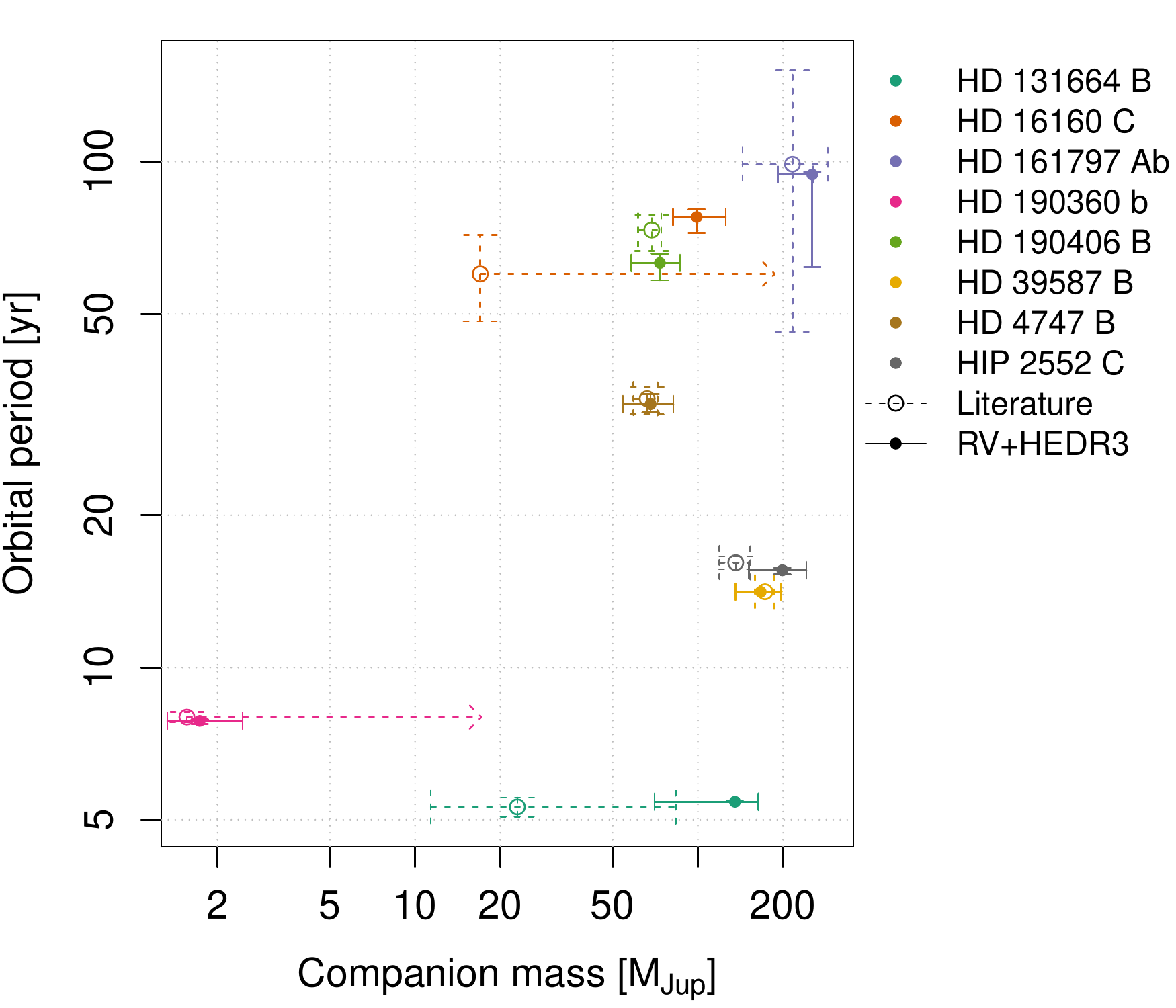}
\caption{Comparison of our estimation of mass and orbital period with
  previous ones for eight nearby systems. The dashed error bars
  represent previous best solutions while the solid ones represent the
  solutions in this work. The mass uncertainty of the primary star is also
considered in the estimation of companion mass. The error bars of mass and period for literature solutions are multiplied by 2.32 to be consistent with the error bars corresponding to 1\% and 99\% quantiles reported in this work. }
  \label{fig:comparison}
\end{figure}

\begin{itemize}
\item HD 131664 (HIP 73408) is a G star with a brown dwarf companion
  detected by \cite{moutou09} through the RV method. Through combined RV and
Hipparcos astrometry data analysis, \cite{sozzetti10} estimated a mass
of 23$_{-5}^{+26}$\mj, an inclination of $55\pm 33$\,deg and a period of $5.3\pm
0.1$\,yr. \cite{reffert11} estimated a mass of
$85.2_{-48.9}^{+54.5}$\mj and an inclination of
$167.1_{-17.8}^{+4.8}$\,deg (the uncertainty corresponds to 3-$\sigma$
confidence region). With both updated RV data and Gaia-Hipparcos offsets, we are able to
constrain the period to be $5.424 \pm 0.004$\,yr, the inclination to be
$170.7 \pm 1.5$, and the mass to be $127.8 \pm 17.9$\mj. Our
  estimated mass is consistent with the value estimated by \cite{reffert11}. Our estimation of orbital period and
other RV-constrained orbital elements are more precise than previous
ones due to the use of updated HARPS data. 

Our estimation of inclination is close to the value in \cite{reffert11} and is quite different than
 the value given by \cite{sozzetti10}. The discrepancy in the estimation of inclination
is probably due to the use of a different convention used by
\cite{sozzetti10}. We use the astrometric convention, which is
the only consistent convention according to \cite{feng19b}. \cite{sozzetti10} probably used the so-called
``observer-independent ascending node'' \citep{feng19b} to define
inclination. These two inclination angles $I_1$ and $I_2$ are related
by $I_1=\pi - I_2$. Thus \cite{sozzetti10}'s value of inclination becomes roughly consistent with ours.

\item HD 16160 (GJ 105) is a triple system containing one K dwarf (A) and two M dwarfs (B and C). The component C has a minimum mass of about 17$M_{\rm Jup}$ according to \cite{golimowski00}. The orbital period is about 60\,yr, the eccentricity is about 0.75, and the semi-major axis is about
  15\,au. Because C is much closer to A than B, C is the main cause of
  the reflex motion of A. Our orbital solution for the C component
  estimates a mass of $102.6\pm 9.4$\mj, orbital period of $76.107 \pm 1.820$\,yr, semi-major axis of $17.0\pm 0.7$\,au and eccentricity of $0.641\pm 0.004$. It is the first time that we have a robust and precise solution for the mass and orbit of GJ 105 C around A. 

\item HD 161797 ($\mu$ Her) has a 2+2 architecture, including $\mu$
    Her Aa and Ab as well as $\mu$ Her B and C. $\mu$ Her Ab is an M
  dwarf with a mass of about 0.32\msun, $P=98.9\pm22.7$\,yr, $a=2.9\pm
  0.3''$, $I=62.82\pm 4.66$\,deg, and $e=0.44\pm 0.06$
  \citep{roberts16}. Our analysis estimates a mass of
  $0.22\pm 0.02$\msun, orbital period of
  $81.610 \pm 8.849$\,yr, eccentricity of $0.386 \pm 0.025$,
  and inclination of $58.8\pm 2.4$\,deg. The difference between the astrometry of Hipparcos and Gaia is unlikley caused by $\mu$ Her B and C because they are about 300\,AU from $\mu$ Her A and thus barely induce any significant nonlinear motion of A over two decades.
Although our optimal value of period differ from \cite{roberts16},
they are marginally consistent with each other due to large
uncertainty in previous estimations. Compared with the Lick RV data
and the 15-year ground-based sub-arcsec astrometric data used by
\cite{roberts16}, the combined Lick and Keck data as well as the long
baseline and sub-mas precision of Hipparcos and Gaia data are perfect
to constrain wide orbit companions such as HD 161797 Ab. Our work demonstrates the ability of combined radial velocity and astrometric analysis in determining stellar mass to 5\% precision. 
  
\item HD 190360 (GJ 777A) is a binary system including a G and M
  dwarf. The M dwarf companion is thousands of au away from the
  primary and thus does not induce significant motion on the primary G
  dwarf over decades. At least two planets are found to orbit the
  primary G dwarf \citep{naef03,vogt05}. According to previous
  studies, HD 190360 b is a Jupiter like planet with a minimum mass of
  $1.56\pm 0.13$\mj, period of $7.98\pm 0.08$\,yr and eccentricity of
  0.313$\pm$0.019 \citep{wright09}. Based on the combined RV and
  astrometry modeling, our solution gives a mass of $1.8\pm 0.2$\mj
  and an inclination of $80.2\pm 23.2$\,deg, the longitude of
  ascending node is $306.6\pm 44.4$. The other parameters are
  constrained to a precision comparable with the ones given in the
  previous analysis of RV data \citep{wright09}. Like $\epsilon$ Indi
  Ab \citep{feng19c}, HD 190360 b is another Jupiter analog which has well determined mass and orbit
based on combined RV and astrometric analysis. HD 190360 b is separated
from HD 190360 A by about 0.25$''$ and is detectable by the
Coronagraph Instrument (CGI) of Roman Space Telescope \citep{spergel13}.

\item HD 190406 (15 Sagittae or GJ 779 or HR7672) is a G star hosting
  a brown dwarf companion discovered by \cite{cumming99} through the
  RV method. A later direct imaging of this system shows that the
  brown dwarf companion probably has a mass of 55-78\mj
  \citep{liu02}. Based on combined analysis of updated imaging
  and RV data, \cite{crepp12} found that HD 190406 b has a
   mass of $68.7_{-3.1}^{+2.4}$\mj, and is on an edge-on orbit with an
   inclination of $97.3_{-0.5}^{+0.4}$\,deg, an eccentricity of
   0.50$_{-0.01}^{+0.01}$, and an orbital period of
   73.3$_{-2.9}^{2.2}$. We use the updated RV data and Gaia-Hipparcos offsets to estimate a
mass of $72.8\pm 6.1$\mj, an inclination of $90.7 \pm 5.0$\,deg, a
period of 61.7$\pm$1.6\,yr, and an eccentricity of
0.462$\pm$0.007. Our estimated orbital period is more than 3-$\sigma$
less than the one given by \cite{crepp12} though we benefit from
nearly double the RV data used by \citep{crepp12} as well as the
Gaia-Hipparcos offsets providing a two-decade baseline. With a more conservative estimation of parameter
uncertainties by considering stellar mass uncertainty, our estimated uncertainty
of companion mass is larger than the one given by \cite{crepp12} while
the precision of other parameters are either comparable with or higher
than those in \cite{crepp12}. 
  
\item HD 39587 ( $\chi^1$ Orionis or GJ 222) is a G0V binary star
  with an orbital period $14.119\pm 0.007$\,yr, an eccentricity of
  $0.451\pm 0.003$ and a mass ratio of $0.15\pm 0.005$
  \citep{han02}. Our solution estimates a mass of $167.2\pm 13.0$\mj, an orbital period of
$14.115\pm 0.005$\,yr, a semi-major axis of
$6.3\pm 0.2$\,au and an eccentricity of
$0.450\pm 0.002$ for HD 39587 B. By including SOPHIE precision RV data as
well as Gaia astrometry, our solution is consistent with previous
solution based on combined analysis of RV and ground-based astrometry
\citep{han02}. As we see in the bottom panels of Fig. \ref{fig:fit1},
the fit to Gaia-Hipparcos offsets is nearly perfect. While SOPHIE RV
data improves the precision of RV model parameters, the Hipparcos and
Gaia catalog data help to constrain all orbital parameters to a
precision as high as the ones achieved through intensive astrometric
observations covering the whole binary orbit \citep{han02}. 

\item HD 4747 (GJ 36) is a K0V spectroscopic binary discovered by \cite{nidever02}. Based on an analysis of the Keck/HIRES radial velocity data, the companion is found to orbit the
primary with a minimum mass of 42.3\mj, period of $18.7\pm 1.8$\,yr and
eccentricity of 0.64$\pm$0.06. The companion is further observed through
direct imaging \citep{crepp16,crepp18} and is characterized as a L/T
transition brown dwarf with a mass of 65.3$_{-3.3}^{+4.4}$\mj,
period of $37.85_{-0.78}^{+0.87}$\,yr, and an eccentricity of
$0.740_{-0.002}^{+0.002}$ based on a combined fit to imaging and
radial velocity data \citep{crepp18}. Through combined analysis of
\cite{crepp18}'s data and the Gaia-Hipparcos proper motion offset,
\cite{brandt19} estimate a mass of $66.2_{-3.0}^{+2.5}$\mj and an
orbital period of $34.0_{-1.0}^{+0.8}$\,yr. Moreover, \cite{peretti19}
derived a dynamical mass of $70\pm 1.6$\mj and an orbital period of
$33.08\pm 0.70$\,yr based on a combined analysis of imaging and RV data. 

Our solution of a planet mass of $68.3\pm 5.9$\mj, an orbital period
of $33.229\pm 0.576$\,yr, and $e=0.732\pm 0.002$, is comparable with
the solutions provided by \cite{crepp18}, \cite{brandt19}, and
\cite{peretti19} but without using direct imaging data. 
Moreover, our value of orbital period is closer to \cite{brandt19}'s
than to \cite{crepp18}'s. This demonstrate again that Gaia and
Hipparcos catalog data is able to constrain companion orbit to a
high precision. 

\item HIP 2552 (V547 Cassiopeiae or GJ 22) is a hierarchical system including at least three components with masses of 0.36, 0.12, 0.18\msun \citep{mccarthy91}. The secondary components (B, C) orbit the primary (A) with semi-major axes of 40 and 5\,au \citep{mccarthy91}. The orbital period for C is 15.4\,yr, and the
inclination is 27\,deg \citep{soderhjelm99}. A study based on the
speckle interferometry estimate $P=16.12\pm0.2$\,yr, $e = 0.18\pm
0.03$, $a = 0.529 \pm 0.005$, $I=46\pm 1$\,deg, the mass of C is $m_{\rm C}
=0.13\pm 0.007$\msun, the mass of A is $m_{\rm A} =0.378_{-0.025}^{+0.028}$\msun
\citep{woitas03}. Based on combined RV and astrometry analysis, we
estimate a mass of $0.19\pm 0.02$\msun, an orbital period of
$15.558\pm 0.097$\,yr, and an eccentricity of $0.134 \pm 0.015$,
constraining the orbit more precisely than previous studies. 
\end{itemize}

\section{Sensitivity tests}\label{sec:sensitivity}
In this section, we compare orbital solutions based on different
  versions of Gaia and Hipparcos data in subsection \ref{sec:solution} and test the
sensitivity of orbital solutions to various systematics in
subsection \ref{sec:effect}.

\subsection{Comparison of orbital solutions}\label{sec:solution}
The orbital solutions based on the combined analyses of Gaia EDR3 and
  RV data are called ``RV+HEDR3'' solutions. To test whether our
method is optimal compared with other methods, we analyze the RV data
in combination with the Hipparcos-Gaia Catalog of Acceleration (HGCA;
\citealt{brandt18,brandt21}). This method assumes that all known
systematics are correctly removed from the proper motions of Gaia
  and Hipparcos. Thus no jitter or offsets are used to model
systematics. This method is frequently used in combined analyses of RV and astrometry data (e.g.,
\citealt{brandt19a} and \citealt{brandt19}).
In these studies, only the proper motion difference between
  Hipparcos and Gaia is used to constrain the companion-induced
  acceleration. Although the Gaia-Hipparcos positional difference of a
  star is used to derive a third proper motion to calibrate the proper motions
  in the two catalogs, the positional difference does not appear in
  the astrometry model or likelihood and thus does not put additional
  constraint on the companion's orbit. By accounting for systematics
  such as frame rotation and light travel time in the Hipparcos and
  Gaia DR2 catalog data, \cite{brandt18} and \cite{brandt21} provide catalogs of
  Hipparcos and Gaia proper motions as well as the third proper
  motion derived from the Hipparcos and Gaia positional
  difference. The DR2-based and EDR3-based HGCA catalogs are called
  ``HGCA2'' and ``HGCA3'', and the solutions based on them are named
  ``RV+HGCA2'' and ``RV+HGCA3'', respectively. To constrain an orbit
  using the HGCA catalogs, we only model the proper motion difference
  (i.e. $\bm{\hat\eta}\equiv [\hat\mu_\alpha,\hat\mu_\delta]$) without
  using jitters and offsets in the astrometry model. Moreover, to test
  whether the RV+HEDR3 soultions shown in Table \ref{tab:edr3} are
  sensitive to the choice of Gaia data releases, we use Gaia DR2
  instead of Gaia EDR3 in the combined analyses of RV and astrometry
  data. The corresponding solutions are labeled ``RV+HDR2''. 

The orbital period and mass of the eight stars for the RV+HDR2,
RV+HGCA2, RV+HEDR3, and RV+HGCA3 solutions are shown in Fig. \ref{fig:4set}. For HD 190406 B, HD 39587
B, HD 4747 B, and HIP 2552 C, the literature solutions provide a good
reference to compare the other three solutions due to their high
precision (see lower panels of Fig. \ref{fig:4set}). For HD 190406 B, there is a $\sim 3\sigma$ tension between
the literature solution and the other three solutions, indicating
unknown systematics either in the direct imaging data used by
\cite{crepp12} or in the Gaia-Hipparcos data used in this work. The
latter is unlikely to be the cause if significant systematics are removed from the HGCA catalogs as claimed by
\cite{brandt18}. For HD 39587 B and HD 4747 B, all solutions constrain
the mass and period to a similar precision. For HIP 2552 C, the mass
error in the RV+HGCA2 solution is about 6 times larger than those in the
RV+HDR2 and RV+HEDR3 solutions.

It is surprising to find that the RV+HEDR3 solutions are not more
  precise than the RV+HDR2 solutions for some targets. There are at least two
  reasons. First, most targets in this work have mass uncertainty of
  $>10$\% (as shown in Table \ref{tab:rv}), making precise estimation
  of companion mass unlikely. The stellar mass errors contribute 7\%,
  67\%, 36\%, 5\%, 82\%, 95\%, 27\%, and 84\% to the uncertainty of
  the masses of HD 131664 B, HD 16160 C, HD 161797 Ab, HD 190360 b, HD
  190406 B, HD 39587 B, HD 4747 B, and HIP 2552 C, respectively.  Second, because
  the five-parameter astrometry model used in Gaia EDR3
  \citep{lindegren21} does not account for companion-induced nonlinear
  motion of a star, the longer baseline of
Gaia EDR3 does not improve the fit much for stars with massive companions. In
particular, the Gaia EDR3 astrometry is more uncertain than the DR2
astrometry for HD 131664, HD 16160, and HIP 2552. Hence it is the
model rather than the data that limits the astrometric precision of Gaia
catalog. The future Gaia
  DR3\footnote{\url{https://www.cosmos.esa.int/web/gaia/release}} will
  identify non-single stars and will thus increase the precision of
  global calibration of Gaia systematics. However, the small nonlinear
  stellar motions caused by exoplanets are unlikely to be identified
  until the final Gaia data release and follow-up analyses of
epoch data \citep{perryman14}. Due to the above reasons, our modeling
of both positional and proper motion difference (RV+HDR2 and RV+HEDR3)
does not significantly improve parameter precision compared with
HGCA-based solutions (RV+HGCA2 and RV+HGCA3) for massive companions in
our sample. However, the mass of the Jupiter analog HD 190360 b in the
RV+HEDR3 solution is 2.2, 7.8, and 3.3 times more precise than the masses in the
RV+HDR2, RV+HGCA2, and RV+HGCA3 solutions, respectively. This demonstrates the
importance of using both position and proper motion of Hipparcos and
Gaia to detect cold Jupiters and constrain their masses.

\begin{figure}
  \centering
  \includegraphics[scale=0.45]{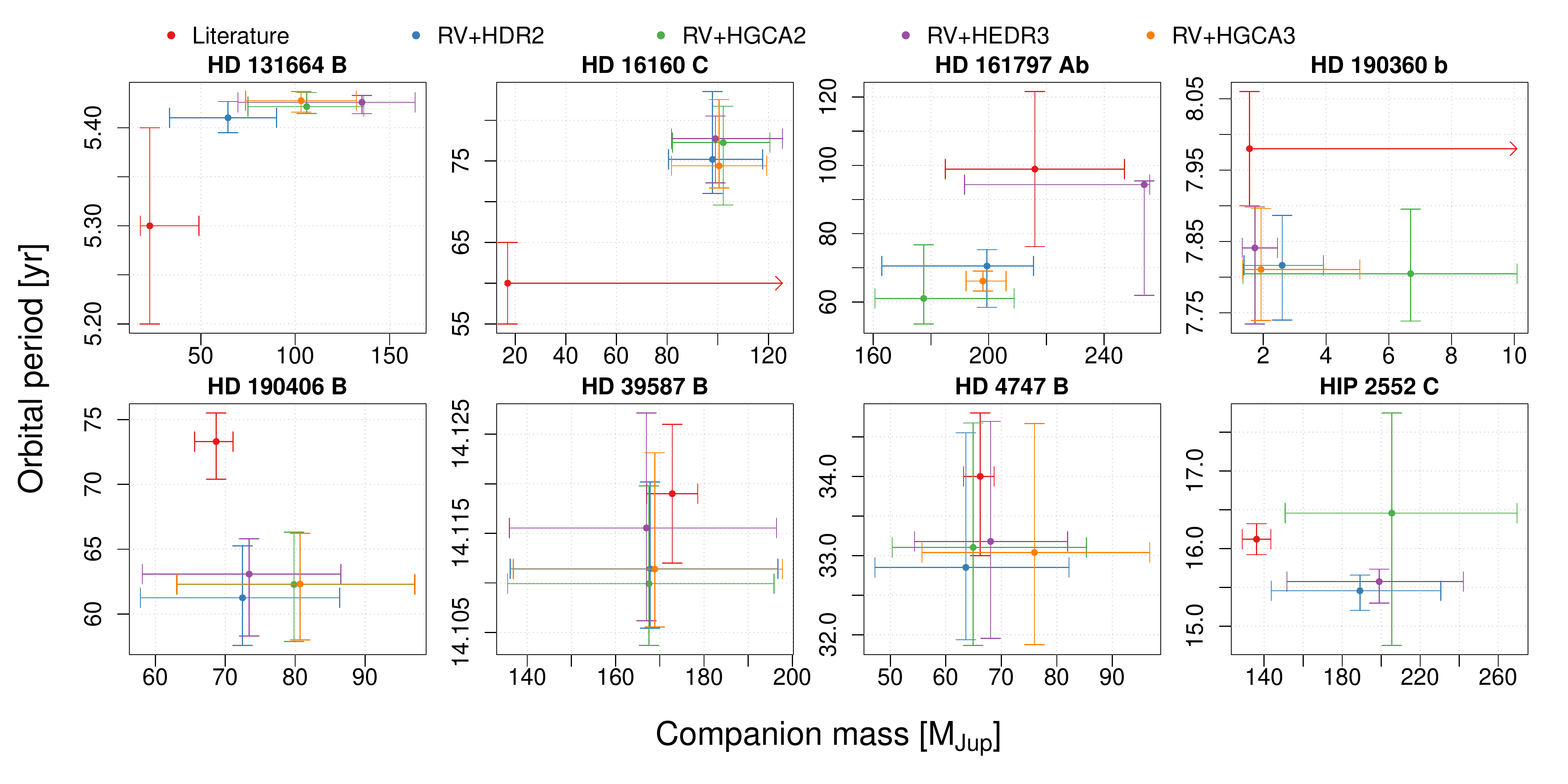}
  \caption{Comparison of the RV+HDR2, RV+HGCA, RV+HEDR3, and
      RV+HGCA3 solutions with the literature solution for each of the
    eight targets. There is no HGCA3 data for HIP 2552 C, and thus only
    three solutions are shown for this target. As in
    Fig. \ref{fig:comparison}, the error bars of the literature
    solutions are adjusted to be consistent with the 1\% and 99\%
    quantiles used in this work. The definition of solutions and various Gaia data sets is introduced in section \ref{sec:solution}.}
  \label{fig:4set}
\end{figure}

\subsection{Sensitivity to various effects}\label{sec:effect}
In our work, instead of calibrating the systematics using a third
  proper motion derived from the positional difference between Hipparcos
  and Gaia, we treat them as unknowns and use jitters and offsets to
  model them. By doing so, we simultaneously fit the signal model and
  noise model to the original positions and proper motions of
  Hipparcos and Gaia catalogs. In this section, we investigate whether various systematics and biases would significantly
influence the RV+HDR2 orbital solutions if the offset and jitter
parameters do not successfully model them. Since RV+HDR2 and RV+HEDR3
are based on the same methodology and have similar precision, such
sensitivity tests are also valid for RV+HEDR3, the default solutions
shown in Table \ref{tab:edr3}.

We first investigate how strongly Gaia-Hipparcos data constrain the
orbits of the companions. Considering that the RV data put strong
constraints on the orbits (see Fig. \ref{fig:fit1} and
\ref{fig:fit2}), we only consider the increase of logarithmic
likelihood ($\Delta$ln$\mathcal{L}$) due to
the addition of astrometric constraint. In other words, we calculate the
likelihoods for the models with and without companions for the astrometric data. The lnBF are shown in Table
\ref{tab:sensitivity}. As seen from Table \ref{tab:sensitivity}, the lnBFs for all
targets are larger than 500, strongly favoring the companion hypotheses. 

Second, considering that the unresolved companions will lead to large
astrometric residuals in Gaia DR2, we expect positive correlation
between the amplitude of stellar reflex motion and
the so-called Renormalised unit weight error (RUWE), a measure of
excess noise suggested by \cite{lindegren18b}. We calculate the
position variation ($\Delta r_{\rm gdr2}$) over the time span of Gaia
DR2 (from 25 July 2014 to 23 May 2016), and find a strong positive
correlation (Pearson's correlation coefficient $r=0.86$) between
$\Delta r_{\rm gdr2}$ and RUWE. Moreover, HIP 2552 has the largest
RUWE while its companion is resolved by Gaia with a designation of
527956488339113600. This indicates that the large RUWE for HIP 2552 is
exactly caused by the perturbation from its companion rather than by the blending of the two
components. Hence RUWE is actually a good indicator of companion-induced acceleration. It should
not be used to assess the quality of DR2 data for nearby stars that might host
massive companions, as also noticed by \cite{kervella19}. 
\begin{table}
  \centering
\small
   \caption{Tabulated results presented for a variety of
     sensitivity tests, with each effect and the measure of its
     significance given in the first column.}
    \label{tab:sensitivity}
     \begin{tabular}{l llll lllr}
       \hline\hline
        & HD131664& HD16160& HD161797& HD190360& HD190406& HD39587&HD4747&HIP2552\\\hline
       Astrometric significance (lnBF) &2328&851&4788&3680&3054&64237&4957&11377\\
       $\Delta r_{\rm gdr2}$ (mas)&2.32&69.92&37.84&0.64 &16.87&14.24&4.97&121.78\\
       RUWE ($\Delta$ln$\mathcal{L}$)  &1.35&1.18&1.38&0.83&0.95&1.03&0.86&4.24\\
       Frame rotation ($\Delta$ln$\mathcal{L}$)&-0.05&-0.07&-0.13&-0.86 &-1.50&-0.25&-3.07&-0.14\\
       Zero parallax offsets ($\Delta$ln$\mathcal{L}$)
        &-0.09&0.56&0.10&0.01&-0.21&-0.64&-0.10&0.06\\
       Timing bias ($\Delta$ln$\mathcal{L}$) &-0.021&0.205&0.005&-0.077&0.029&-2.989&-0.004&-0.028\\
       Companion flux ($1000 \eta$) &0.0041&1.6&5.6&$5.8\times 10^{-36}$&0.068&2.9&0.098&0\\
       \hline
       \end{tabular}
\end{table}

Third, we assess the influence of frame rotation, zero parallax
offsets, and timing bias on the significance
of astrometric signal. Following \cite{lindegren18} and
\cite{kervella19}, we use the rotation parameters,
$w_x=-0.086\pm0.025$\,mas\,$s^{-1}$,
$w_y=-0.144\pm0.025$\,mas\,$s^{-1}$,
$w_x=-0.037\pm0.025$\,mas\,$s^{-1}$ to correct the Gaia DR2 proper
motions. After this frame rotation correction, we find that the
logarithmic likelihoods of the solutions decrease slightly probably because the
offset parameters are already optimized to account for frame
rotation and the addition of {\it a priori} frame rotation makes the
fit slightly worse. Since the frame rotations of Hipparcos and Gaia
DR2 relative to the International Celestial Reference System (ICRS) have similar amplitudes \citep{brandt18}, we conclude that our results are not
sensitive to the correction of frame rotation. We also correct the
0.029\,mas zero parallax offset and find that the variation in
logarithmic likelihood is less than 1 for all targets. Moreover, we
correct for the distance change due to light travel time \citep{kervella19} and find less than 0.01 variation in
logarithmic likelihood.  We also correct the timing bias due to
the difference between the central and reference epochs for Gaia and
Hipparcos. We only find small changes in ln$\mathcal{L}$  and the maximum
$\Delta$ln$\mathcal{L}$ is negative for HD4747, indicating that the offset parameters
are enough to account for the timing bias. 

Finally, we calculate the offset of the photon center of a system
relative to the primary star. This offset is orbital
dependent and thus will bias orbital parameters, especially the
companion mass. Hence we calculate the fractional
offset ($\eta$) of photon center ($r-\delta r$) relative to the
primary star ($r$) in the barycentric reference frame. By comparing
$\eta$ with the fractional uncertainty of orbital parameters, we can
test the sensitivity of the orbital solutions to the companion flux. 
For an approximate calculation of the companion flux, we first derive
luminosity from mass using $L\propto M^{3.5}$. We then
calculate the effective temperatures of HD 131664 B, HD190360 b, HD 190406 B, and HD 4747 B using the cooling
models introduced by \cite{phillips20} and \cite{marley18}. The
procedure will be introduced in the following section. The
effective temperatures of HD 161797 and HD 190360 are respectively given by
\cite{grundahl17} and \cite{ligi16}. The temperatures of other stars
are either given by Gaia DR2 or determined by using table 5 of
\cite{pecaut13} and its updated version at \url{http://www.pas.rochester.edu/~emamajek/EEM_dwarf_UBVIJHK_colors_Teff.txt}. These quantities
are used to determine the fractional offset $\eta$, which is
\begin{equation}
\eta=\frac{\delta r}{r} \approx \left(\frac{M_p
  }{M_\star}\right)^{2.5}\cdot \frac{f(T_p)}{f(T_\star)}~,
\label{eqn:eta}
\end{equation}
where
\begin{align}
  f(T_p)&=\frac{\int B(T_p,\lambda)F(\lambda)d\lambda}{\int
          B(T_p,\lambda) d\lambda}~,\\
  f(T_\star)&=\frac{\int B(T_\star,\lambda)F(\lambda)d\lambda}{\int
    B(T_\star,\lambda) d\lambda}
\end{align}
are the fraction of the flux passing the Gaia G passband or the Hipparcos passband relative to the total flux,
$T_p$ and $T_\star$ are respectively the effective temperatures of a companion and
its primary star, 
\begin{equation}
  B(T,\lambda)= \frac{8\pi hc}{\lambda^5(e^{(hc/\lambda\kappa T)}-1)}
\end{equation}
is the Planck's law of blackbody radiation, $h$ is the planck
constant, $c$ is the speed of light, $\kappa$ is the Boltzman
constant. The transmission data ($F(\lambda)$) for the G passband is given by \cite{evans18}
while the Hipparcos passband data is from \cite{bessell00}. By applying Equation \ref{eqn:eta} to the eight target stars, we
estimate and show $\eta$ for the Gaia G passband. Because $\eta$ for the Hipparcos passband is less than that
for the G passband for all stars, we do not show them in the table. The companion of HIP 2552 was
resolved, and thus contribute zero flux to the G band flux measured
for its primary. However, Hipparcos did not resolve the pair and the corresponding $\eta$ is about 1\%. As
shown in Table \ref{tab:sensitivity}, all stars have no more than 1\%
fractional offset caused by companion flux. Because the uncertainty
caused by companion flux is far less than the uncertainty of the
orbital parameters, it does not play a significant role in this work.  

We emphasize that the above tests assume that the offset and jitter
parameters did not model all systematics in the Hipparcos and Gaia
data. Even under this unlikely assumption, we find that
the correction of systematics do not have significant impact on the
orbital solutions. Therefore, we conclude that our orbital analyses
are robust to the systematics mentioned in previous studies. 

\section{Characteristics of Jupiter analog and brown
  dwarfs}\label{sec:BD}
Among the eight low mass companions, HD 190360 b is a Jupiter analog while HD
  190406 and HD 4747 B are brown dwarfs. In this section, we use
  evolution models to estimate their effective temperatures or ages for direct imaging and
  follow-up studies. We use the recently developed \texttt{ATMO 2020} model
\citep{phillips20} to analyze the brown dwarfs and giant exoplanets
{\rm based on the RV+HEDR3 solutions}. Since HD 190360 b is not directly imaged, we assign ages from
the literature to them in order to determine their effective temperature
and potential atmospheric composition. On the other hand, HD 190406 B
and HD 4747 B are directly imaged brown dwarfs and thus are suitable
targets for independent age constrants. The other targets in this work
are low-mass M dwarfs and thus do not follow the brown dwarf
cooling tracks. Hence we will focus our investigation on the four
targets that we mention above.

The age of HD 190360 is quite uncertain \citep{ligi16} and thus we adopt a relatively conservative estimation
of 2.8-9.4\,Gyr by \cite{bensby14}. The effective temperature of HD
4747 B (1700$\pm$100\,K) is derived from photometry using the \cite{baraffe03}
evolutionary models by \cite{crepp16}. The temperature of HD 190406 B
(1510-1850\,K) is derived from photometry using the \cite{burgasser03}
model by \cite{liu02}. 

We show the \texttt{ATMO 2020} evolution tracks as well as the
effective temperature and mass of the four targets in
Fig. \ref{fig:evolution}. For HD 190406 B and HD 4747 B, we estimate an age of $>1.5$\,Gyr and
1-3\,Gyr, respectively. Our independent estimation of the age of HD 190406 B is consistent with the value of 1.8$\pm$0.4\,Gyr derived from
lithium abundance of HD 190406 by \cite{aguilera18}. However, our
estimated age of HD 4747 B is younger than the
$10.74_{-6.87}^{+6.75}$\,Gyr estimated by \cite{wood19} using stellar
evolution tracks or isochrones based on three different stellar models
but is consistent with the value of 0.9-3.7\,Gyr
adopted by \cite{peretti19} based on the age-log$R_{\rm HK}$
calibration introduced by \cite{mamajek08}. This indicates a potential
discrepancy between stellar and substellar evolution models though
more examples are needed. 

\begin{figure}
  \centering
  \includegraphics[scale=0.8]{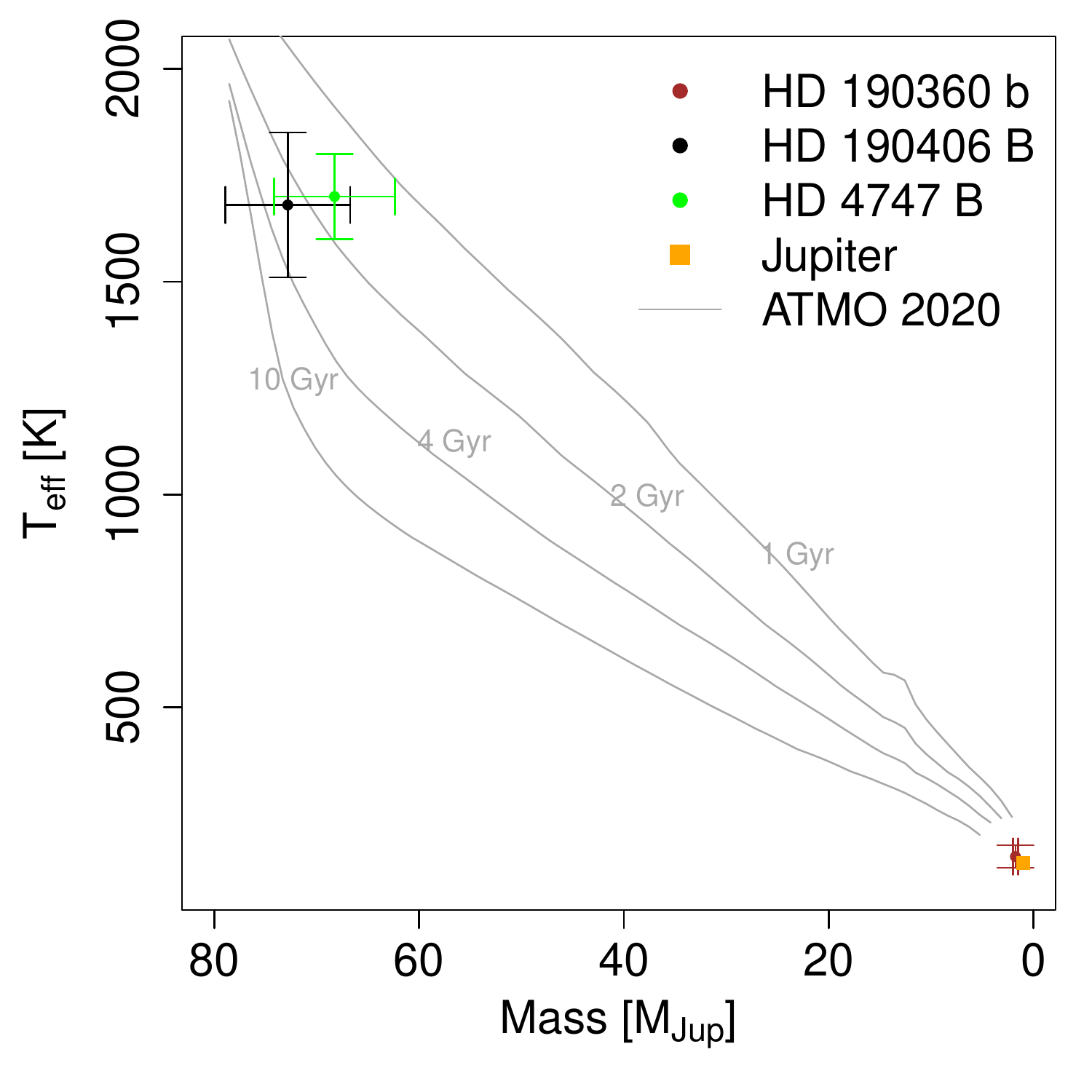}
  \caption{Evolution model for HD 190360 b, HD 190406 B and HD 4747 B. The grey isochrones represent the
    \texttt{ATMO 2000} cooling tracks of brown dwarfs and giant
    planets. The error bars of the effective temperatures of HD 190406
    B and HD 4747 B are adopted from literature. The temperature of
     HD 190360 b is derived from the evolution tracks
    assuming literature ages. Jupiter in our solar system is also shown as an orange square for reference. }
  \label{fig:evolution}
\end{figure}

For HD 190360 b, the uncertainty of the dynamical mass
and the lack of photometry mean the effective temperature is only
loosely constrained by observations. Since the \texttt{ATMO 2020}
cooling tracks do not extend to objects with $T_{\rm eff}<200$\,K, we
apply the \texttt{Sonora 2018} model developed by \cite{marley18} and find a temperature of 123-176 K for HD 190360 b, similar to the effective
temperature of our Jupiter (134\,K; \citealt{aumann69}). Considering
that HD 190360 is a Sun-like star and possibly has the same age as the
Sun, HD 190360 b is likely to have a Jupiter-like atmosphere mainly
made of molecular hydrogen and helium. Hence HD 190360 b becomes a new
member of a small sample of nearby Jupiter analogs with well
constrained orbits, including eps Indi A b \citep{feng19c} and
HAT-P-11 c \citep{xuan20}. HD 190360 b is about 0.25$''$ away from
its host, making it a good candidate for direct imaging by the CGI
instrument of the Roman Space Telescope \citep{bailey18,feng20d}.

\section{Discussion and conclusion}\label{sec:conclusion}
We constrain the mass and orbital elements for eight low mass companions
to a high precision based on combined analysis of the RV data and the
proper motion and positional difference between Hipparcos and Gaia
catalog data. In particular, the consistency between our orbital
solutions and the previous best solutions for HD 4747 B and HIP 2552 C at a high
precision level justifies the use of Gaia-Hipparcos offsets in both position
and proper motion to constrain wide companions. This approach was also
successfully used to detect $\epsilon$ Indi Ab
\citep{feng19c}. We also use the \texttt{ATMO 2020} models to derive the ages of HD 190406 B and HD 4747
B, consistent with the most recent age estimates. The age of HD 190406
derived from stellar evolution models is found to be older than the age derived
using brown dwarf cooling models or age-log$R_{\rm HK}$
calibrations. More such examples need to be found to investigate the
causes of such discrepancy. 

Our application of the combined analysis to the data for HD 16160, HD
161797 A, HD 190360, and HD 39587 significantly improves the
constraint of their companions' orbits. For HD 16160 C and HD 190360 b, the Gaia-Hipparcos data enables the
degeneracy between mass and inclination of RV-only analysis to be
solved for the first time. The application of Gaia
and Hipparcos data to analyses of a large sample of low mass
companions will offer an efficient way to find the most promising
targets for follow-up imaging observations which in the end can offer
the strongest constraints on evolution and atmospheric models for
brown dwarfs and giant planets.

In particular, with a mass of $1.8\pm 0.2~M_{\rm Jup}$, HD 190360 b becomes the smallest Jupiter analog with well
determined mass and orbit based on combined analysis of RV and
astrometry data. It is separated from its host star by 0.25$''$
and is thus suitable for direct imaging by CGI of Roman Space Telescope. Like $\epsilon$ Indi Ab, it belongs to a very
small sample of cold Jupiters with well constrained orbits. Although
$\epsilon$ Indi Ab may be observable by MIRI/JWST or next-generation facilities \citep{matthews21,pathak21}, these
targets are typically too faint for current direct imaging facilities. Although
the microlensing method is sensitive to Jupiter analogs, microlensing
events are rare, likely to be distant, hard to follow-up and do not
fully constrain planetary orbits. Therefore the astrometric survey
currently conducted by Gaia is a perfect resource for a thorough
understanding of the occurrence rate, formation and evolution of
Jupiter analogs \citep{perryman97}. In particular, a large sample of
multiple-planet systems with well-characterized cold Jupiters are crucial to test whether there is a positive correlation
between the occurrence rates of inner super-Earths and outer giant
planets \citep{schlecker20}.

Compared with the previous use of only proper motion difference
between Gaia and Hipparcos and/or a proper motion derived from
  Gaia-Hipparcos difference in position (e.g., \citealt{kervella19}
and \citealt{brandt18}), our modeling of both proper motion and position offsets
in a robust way allows us to constrain orbits of cold Jupiters to a higher
precision. Although this combined RV and astrometry approach is promising, its precision is limited by a lack
of accurately determined masses for the primary stars and the lack of Gaia epoch data. For giant companions
around single stars that have not been imaged, we rely on other methods such the
mass-luminosity relation for mass determination. The
error on the stellar mass is added quadratically to the error of
companion mass and thus limits the precision of giant planet
characterization. Therefore we need high precision methods such as
asteroseismology \citep{epstein14} to independently determine the
stellar mass. A synergy of high cadence and high precision photometry,
RV, direct imaging and astrometry data is essential to fully and
accurately characterize a large sample of circumstellar brown dwarfs
and giant planets for the study of planet formation. 

\section*{Acknowledgements}
We are very grateful to the anonymous referee for their many
insightful comments which greatly improved this work. This work has made use of data from the European Space Agency (ESA)
mission {\it Gaia} (https://www.cosmos.esa.int/gaia), processed by the
{\it Gaia} Data Processing and Analysis Consortium (DPAC,
https://www.cosmos.esa.int/web/gaia/dpac/consortium). Funding for the
DPAC has been provided by national institutions, in particular the
institutions participating in the {\it Gaia} Multilateral
Agreement. This work is also based on observations collected at
  the European Southern Observatory under ESO programmes 072.C-0488,
  089.C-0732, 090.C-0421, 091.C-0034, 093.C-0409, 099.C-0458,
  0103.C-0432, 072.C-0096, 073.D-0038, 074.D-0131, 075.D-0194,
  076.D-0130, 078.D-0071, 079.D-0075, 080.D-0086, 081.D-0065. Part of
  this research was carried out at the Jet Propulsion Laboratory,
  California Institute of Technology, under a contract with the
  National Aeronautics and Space Administration (NASA).

\section*{Data availability}
The data underlying this article are available in the article and in its online supplementary material.

\bibliographystyle{mnras}
\bibliography{nm}

\end{document}